\documentclass[ArXiv,AXISWP,accept,moreauthors,pdftex,10pt,letterpaper]{Definitions/axis} 

\newcommand{\Msun}{\,{\rm M_\odot}}

\newcommand{\Mblack}{M_\bullet}

\firstpage{1} 
\pubvolume{xx}
\issuenum{1}
\articlenumber{5}
\pubyear{2023}
\copyrightyear{2023}
\Title{Detecting Wandering Intermediate-Mass Black Holes with AXIS in the Milky Way and Local Massive Galaxies}

\Author{Fabio Pacucci$^{1, 2\dagger}$, Bryan Seepaul$^{1}$, Yueying Ni$^{1}$, Nico Cappelluti $^{3}$, and Adi Foord $^{4}$}

\AuthorNames{Fabio Pacucci, Bryan Seepaul, Yueying Ni, Nico Cappelluti, Adi Foord}

\address{%
$^{1}$ \quad Center for Astrophysics | Harvard \& Smithsonian, 60 Garden St., Cambridge MA, 02138 \\
$^{2}$ \quad Black Hole Initiative at Harvard University, 20 Garden St., Cambridge MA, 02138 \\
$^{3}$ \quad University of Miami \\
$^{4}$ \quad Department of Physics, University of Maryland Baltimore County, 1000 Hilltop Cir, Baltimore, MD 21250, USA}

\firstnote{Email: fabio.pacucci@cfa.harvard.edu} 
\abstract{This white paper explores the detectability of intermediate-mass black holes (IMBHs) wandering in the Milky Way (MW) and massive local galaxies, with a particular emphasis on the role of AXIS. IMBHs, ranging within $10^{3-6} \Msun$, are commonly found at the centers of dwarf galaxies and may exist, yet undiscovered, in the MW. 
By using model spectra for advection-dominated accretion flows (ADAFs), we calculated the expected fluxes emitted by a population of wandering IMBHs with a mass of $10^5 \Msun$ in various MW environments and extrapolated our results to massive local galaxies.
Around $40\%$ of the potential population of wandering IMBHs in the MW can be detected in an AXIS deep field. We proposed criteria to aid in selecting IMBH candidates using already available optical surveys.
We also showed that IMBHs wandering in $>200$ galaxies within $10$ Mpc can be easily detected with AXIS when passing within dense galactic environments (e.g., molecular clouds and cold neutral medium). 
In summary, we highlighted the potential X-ray detectability of wandering IMBHs in local galaxies and provided insights for guiding future surveys.
Detecting wandering IMBHs is crucial for understanding their demographics, evolution, and the merging history of galaxies.
\emph{This White Paper is part of a series commissioned for the AXIS Probe Concept Mission; additional AXIS White Papers can be found at the  \href{http://axis.astro.umd.edu/}{AXIS website} with a mission overview \href{https://arxiv.org/abs/2311.00780}{here}}.
}

\begin{document}
\tableofcontents
\listoffigures
%%%%%%%%%%%%%%%%%%%%%%%%%%%%%%%%%%%%%%%%%%

\section{Introduction}
Many of the black holes (BHs) observed thus far are accreting at or near the Eddington rate $\dot{M}_{\rm Edd} \approx 1.4\times 10^{18} \Mblack \, \rm g \, s^{-1}$, where $\Mblack$ is the mass of the compact object in solar masses. In this limiting case, the outward acceleration on a test particle resulting from radiation pressure is balanced by the inward gravitational acceleration. 
Notably, this is the case for high-luminosity quasars, characterized by super-massive BHs with masses $\Mblack > 10^6 \Msun$. In the conventional $\alpha$-disk model \citep{Shakura_Sunyaev_1973, Novikov_Thorne}, $\sim 10\%$ of the rest-mass energy ($Mc^2$) of the infalling material is radiated away \citep{Narayan_2005}.

This standard picture of accretion has been widely tested, especially in the high-z Universe \citep{Fan_2022}, where the large availability of gas makes accretion at the Eddington rate feasible \citep{Power_2010}. 
However, the radiative efficiency, $\epsilon$, can significantly deviate from the typical $10\%$ value, both for strongly super-Eddington ($\dot{M} \gg \dot{M}_{\rm Edd}$) and sub-Eddington ($\dot{M} \ll \dot{M}_{\rm Edd}$) accretion rates \citep{Begelman_1978, Paczynski_1982, Abramowicz_1988, Volonteri_2005, Sadowski_2009}.

Below $1\%$ of the Eddington rate, the advection-dominated accretion flow (ADAF) regime is entered \citep{Narayan_1994, Narayan_1995, Abramowicz_1995, Narayan_2008, Yuan_Narayan_2014}. BHs accreting in ADAF mode exhibit radiative efficiencies several orders of magnitude lower than the typical $\sim 10\%$ value. Given the rarity of conditions supporting large accretion rates in the local Universe, it is likely that a substantial fraction of BHs accretes in the ADAF mode, e.g., the super-massive BH at the center of the MW \citep{Yuan_2003}. Similarly, a putative population of intermediate-mass BHs (IMBHs) wandering in galaxies would also accrete in ADAF mode.

IMBHs are a bridge between stellar mass and super-massive objects and have masses in the range $10^3 \Msun < \Mblack < 10^6 \Msun$, although the definition greatly varies depending on the sub-field of interest. Central IMBHs have been extensively detected in dwarf galaxies according to scaling relations \citep{Kormendy_Ho_2013}, and have active fractions from $\sim 5\%$ to $22\%$ \citep{Greene_2020_IMBH, Pacucci_2021_active}. Some of these central black holes in local dwarf galaxies, and up to $z \sim 0.9$, are found to be significantly overmassive with respect to the stellar content of their hosts, in violation of the scaling relations \citep{Mezcua_2023}. Particularly overmassive black holes are now systematically found in the high-$z$ Universe \citep{Pacucci_2023} by JWST. The redshift evolution of these populations of black holes and the role that wandering black holes played in their formation (see, e.g., \citep{DiMatteo_2022}) is still unclear.
For these reasons, recent studies are focusing on investigating the existence of wandering IMBHs in the MW and massive galaxies and their orbital and radiative properties \citep{Ricarte_2021_IMBH, Weller_2022, Weller_2023, DiMatteo_2022}.

IMBHs potentially wandering in the MW could have formed (i) in situ and (ii) ex-situ. In situ (i.e., within the galaxy) formation channels include direct collapse of high-mass quasi-stars \citep{Volonteri_2010_IMBH, Schleicher_2013}, super-Eddington accretion onto stellar-mass BHs \citep{Ryu_2016}, runaway mergers in dense globular stellar clusters \citep{PZ_2002, Gurkan_2004, Shi_2021, Gonzalez_2021, Fragione_2022}, and supra-exponential accretion on seed black holes in the early universe \citep{Alexander_Natarajan_2014, Natarajan_2021}. The ex-situ channel forms the wandering black holes through tidal disruption of satellite/dwarf galaxies when merged into larger halos. \citep{Governato_1994, Volonteri_2003, Oleari_2009, Greene_2021_wandering}.

These wandering IMBHs accrete from the interstellar medium (ISM) at low rates $\dot{M} \ll \dot{M}_{\rm Edd}$, resulting in electromagnetic signatures typical of ADAF accretion mode. 
A recent study \citep{Seepaul_2022} modeled the accretion and radiation properties of putative IMBHs wandering in the MW, of $10^5 \Msun$ in mass, using five realistic ISM environments \citep{Ferriere_2001}: molecular clouds (MCs), cold neutral medium (CNM), hot neutral medium (HNM), warm ionized medium (WIM), and hot ionized medium (HIM). 
MC is the densest environment, with typical gas number densities of $10^{2} - 10^{4} \, \mathrm{cm^{-3}}$, while HIM is the most rarified environment, with typical gas number densities of $10^{-3} \, \mathrm{cm^{-3}}$. All results presented here consider the volume fractions of the different environments considered. MC is the most uncommon environment, occupying only $\sim 0.05\%$ of the volume of the MW, while HIM is the most common, with a volume occupation fraction of $\sim 47\%$: almost half of the entire volume \citep{Seepaul_2022}.
The mass of the perturbing black hole was chosen as the typical mass of IMBHs detected in the nuclei of dwarfs \citep{Greene_2020_IMBH}.
The accretion rate onto the IMBH was estimated using a Bondi rate, adequately adjusted to account for outflows and convection \citep{Igumenshchev_2003, Proga_2003, Perna_2003, Seepaul_2022}.  

This white paper first summarizes the result presented in \cite{Seepaul_2022}, focusing on the X-ray properties of wandering IMBHs in the MW. Then, it expands on the contribution that AXIS \citep{AXIS_2023} could provide to detect these sources. Lastly, it predicts the observability of IMBHs wandering in local galaxies. 
 
%%%%%%%%%%%%%%%%%%%%%%%%%%%%%%%%%%%%%%%%%%
\section{Detecting Wandering IMBHs in the X-rays with AXIS}
\vspace{-10pt}
\subsection{Accretion rates and spectral energy distributions}

\begin{figure}
\centering
\includegraphics[width=0.46\textwidth]{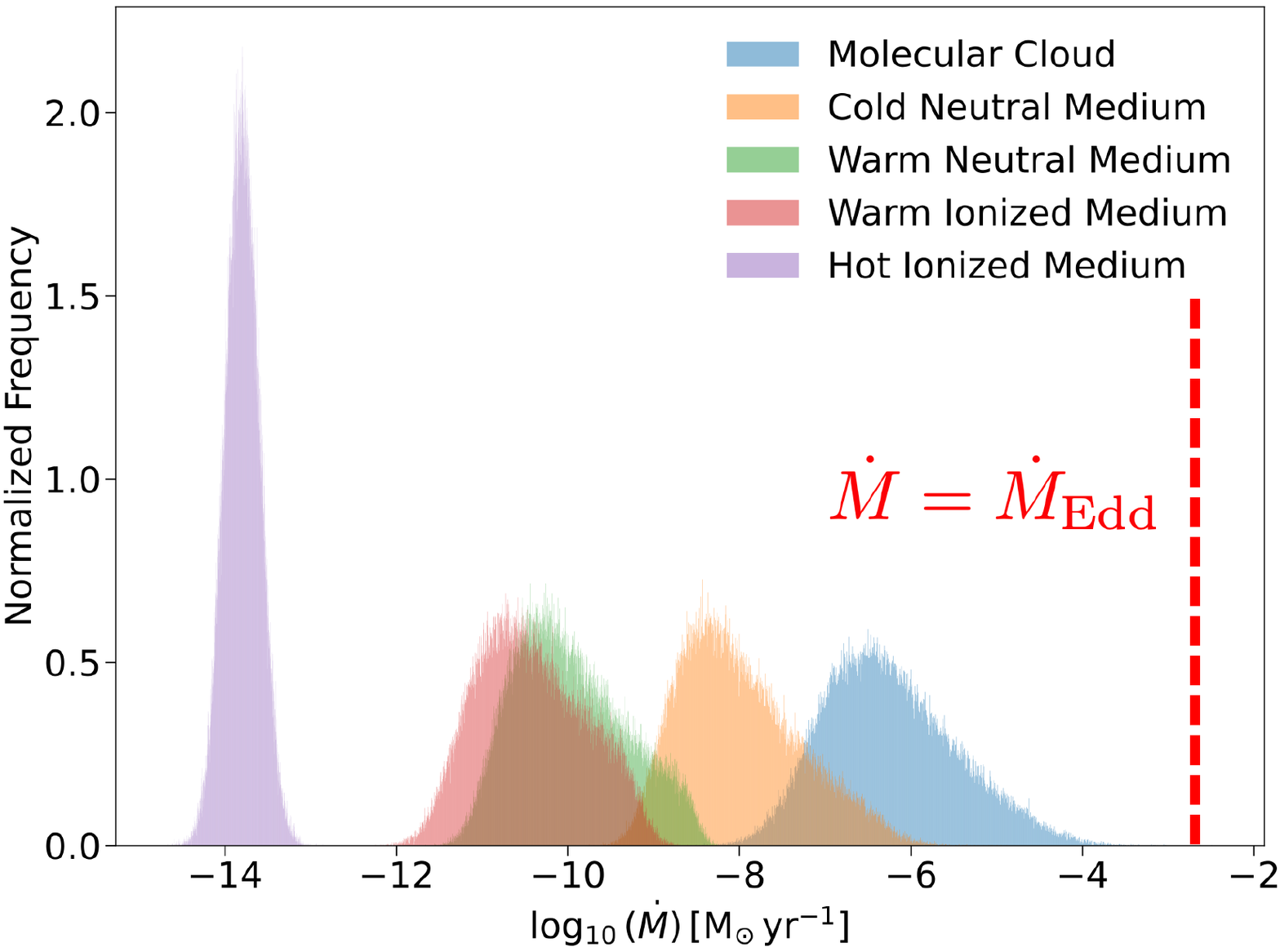}
\includegraphics[width=0.49\textwidth]{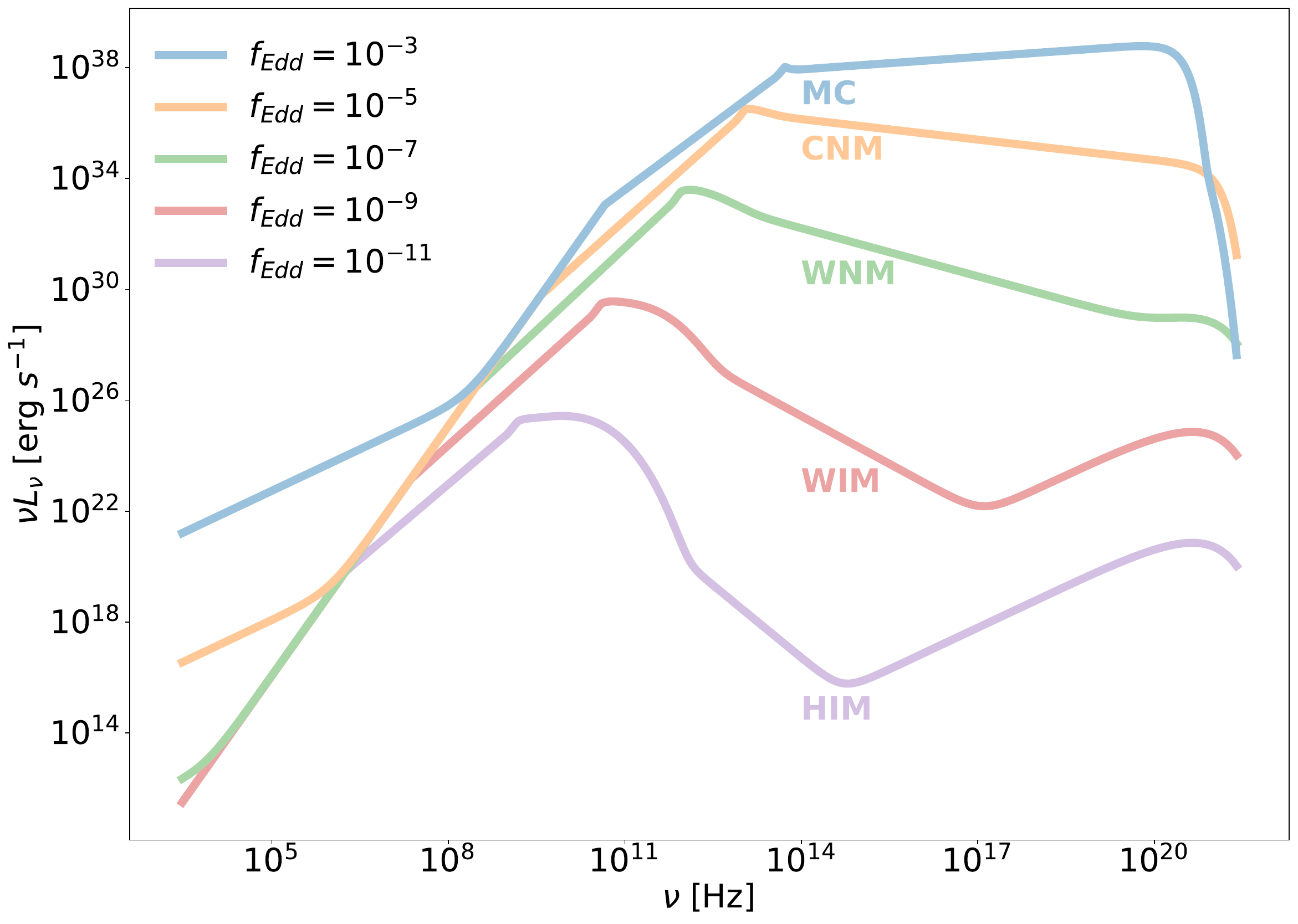}
\caption[Distribution of Typical Accretion Rates and SEDs for the Five ISM Environments Studied]{\textbf{Left panel:} distribution of accretion rates, categorized in the five ISM environments investigated. All rates are strongly sub-Eddington. \textbf{Right panel:} collection of SEDs for five values of the accretion rate representative of each ISM environment.}
\label{fig:accretion_seds}
\end{figure}

The left panel of Fig. \ref{fig:accretion_seds} shows the distribution of accretion rates predicted for a $10^5 \Msun$ IMBH wandering in typical ISM environments of the MW. 
The accretion rates range between $10^{-14}$ and $10^{-4} \mathrm{\Msun \, yr^{-1}}$, hence they span $\sim 10$ orders of magnitude. MC and CNM environments show the highest accretion rates because they are the densest; as such, they offer the best chance for an X-ray detection of IMBHs in the MW.

As a reference, the Eddington rate for a $10^5\Msun$ IMBH is $\dot{M}_{\rm Edd} \approx 2 \times 10^{-3} \mathrm{\Msun \, yr^{-1}}$: all accretion rates predicted are strongly sub-Eddington. The resulting spectral energy distributions (SED), typical for the five ISM environments, are shown in the right panel of Fig. \ref{fig:accretion_seds}, with Eddington ratios (i.e., the actual accretion rate normalized to the Eddington rate) ranging from $10^{-11}$ to $10^{-3}$. The SEDs were calculated using a code designed specifically for ADAF mode accretion \citep{Pesce_2021}. The SED peak shifts to higher frequencies with increasing accretion rates.

\subsection{X-ray observability and selection criteria: the role of AXIS}
\label{subsec:AXIS}

\begin{figure}[H]
\centering
\includegraphics[width=0.47\textwidth]{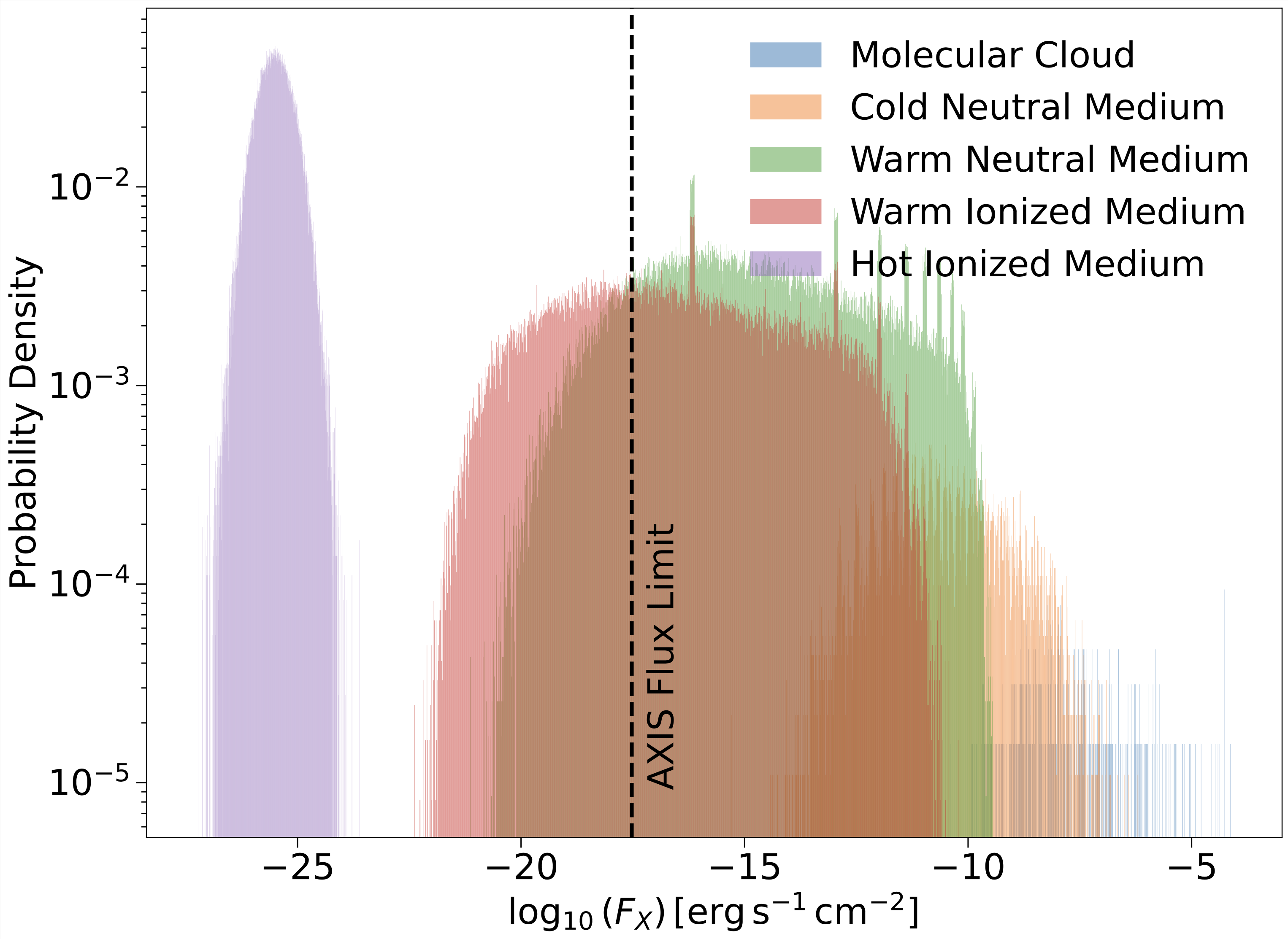}
\includegraphics[width=0.45\textwidth]{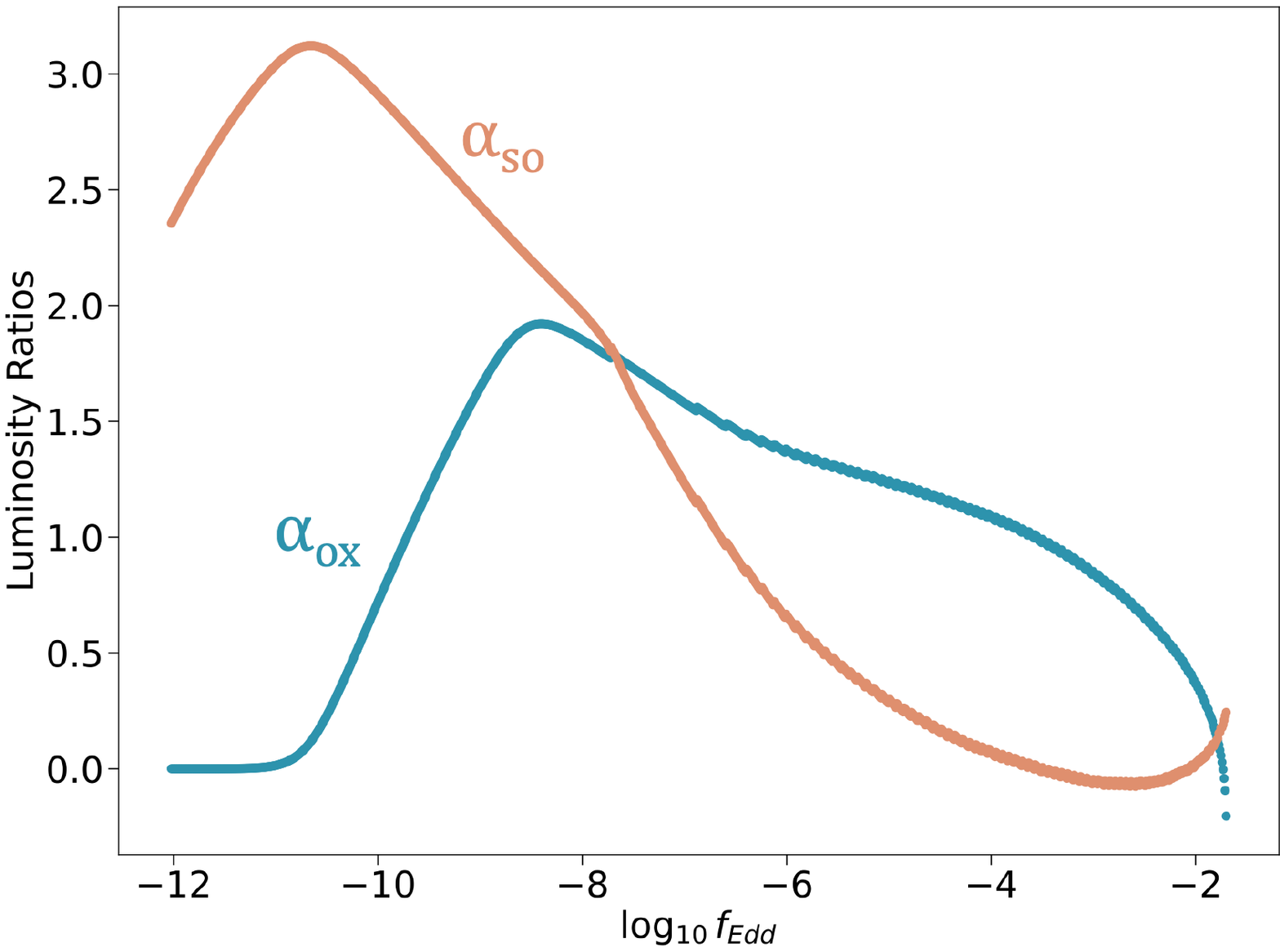}
\caption[Distribution of Typical X-ray Fluxes and Luminosity Ratios]{\textbf{Left panel:} Probability density of the X-ray fluxes produced by IMBHs passing through the five ISM environments considered. The AXIS flux limit is indicated. \textbf{Right panel:} Luminosity ratios (optical to X-ray, and sub-mm to optical) for selecting IMBH candidates in multi-wavelength surveys.}
\label{fig:fluxes_selection}
\end{figure}

The left panel of Fig. \ref{fig:fluxes_selection} shows the resulting (volume-weighted) X-ray flux distribution. These results suggest that AXIS, in its proposed deep survey \citep{Marchesi_2020, Mushotzky_2019} with a flux limit $\sim 3 \times 10^{-18} \rm \, erg \, s^{-1} \, cm^{-2}$ and an area of $0.1 \, \rm deg^{-2}$, will detect a fraction, in number, of $\sim 38\%$ of wandering IMBHs in the MW, assuming a uniform sampling of the region occupied by the Galaxy \citep{AXIS_2023}.

To aid in the task of selecting IMBH candidates, \citep{Seepaul_2022} proposed essential selection criteria to be used in photometric surveys. The right panel of Fig. \ref{fig:fluxes_selection} shows two luminosity ratios calculated as a function of the Eddington ratio of the IMBH: (i) the X-ray to optical/UV ratio represented by the standard $\alpha_{\rm ox}$ parameter \citep{Lusso_2010}, and (ii) the optical/UV to sub-mm ratio (see \cite{Seepaul_2022} for their definition).
A combination of X-ray, optical, and sub-mm observations can sift out potential candidates and uniquely determine the accretion rate onto wandering IMBHs.

Predicting the number of IMBHs detectable by AXIS in a deep Galactic survey is challenging because the total number of such sources is unknown. The MW has encountered $\sim 15 \pm 3$ galaxies with stellar mass $>10^7 \Msun$ during its cosmic evolution \citep{Kruijssen_2020}. Such galaxies could have hosted IMBHs that are massive enough to be detected in AXIS searches. Therefore, assuming an expected number of $\sim 10$ IMBHs, \cite{Weller_2022} showed that these objects are more likely to wander in the innermost $\sim 1$ kpc of the MW. 
However, it is informative to compare AXIS capabilities with other facilities currently operational. Table \ref{table} shows that AXIS would allow for a significant improvement of at least $40\%$ over current facilities, thanks not only to its extraordinary sensitivity but also to its wide field of view.

\begin{table}[H]
\caption{Volume-weighted detectability of wandering IMBHs of $10^5 \Msun$ in the MW by AXIS, Chandra, and eROSITA. The detectability indicates the percent of the total number of objects that are detectable by a given instrument.}
\centering
\begin{tabular}{ccc}
\toprule
\textbf{X-ray telescope}	& \textbf{Flux limit $ \rm [erg \, s^{-1} \, cm^{-2}]$}	& \textbf{Detectability}\\
\midrule
AXIS		& $3.0\times10^{-18}$			& $38\%$\\
Chandra		& $2.0\times10^{-16}$			& $27\%$\\
eRosita		& $2.0\times10^{-14}$			& $13\%$\\
\bottomrule
\end{tabular}
\label{table}
\end{table}

\subsection{Extending the search to local galaxies}

The left panel of Fig. \ref{fig:fluxes_selection} shows that the passage of a $10^5 \Msun$ IMBH generates the highest X-ray fluxes within MCs and CNM environments. 
Galaxies in the same mass category of the MW share a similar environmental composition.
Hence, we investigate the fluxes that the passage of equally massive IMBHs would produce in nearby galaxies.

\begin{figure}[H]
\centering
\includegraphics[width=0.75\textwidth]{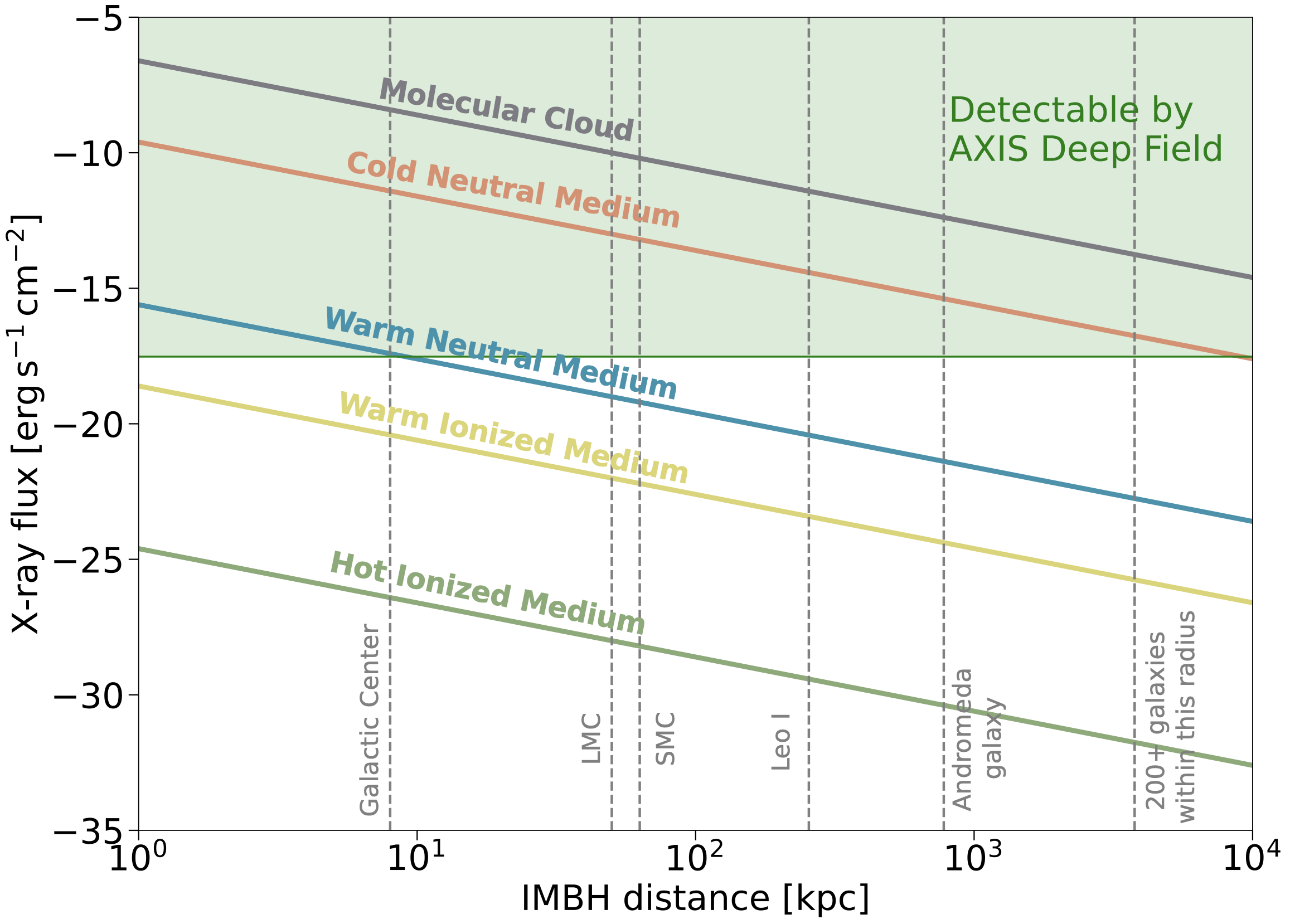}
\caption[Typical X-Ray Fluxes as a Function of Distance Within 10 Mpc]{X-ray fluxes ($0.2-10$ keV) generated by the passage of a $10^5 \Msun$ IMBH as a function of its distance. The green-shaded area is detectable by an AXIS deep field. Electromagnetic signatures of the passage of an IMBH can be detected in MC and CNM in hundreds of galaxies within 10 Mpc.}
\label{fig:extragalactic}
\end{figure}

In Fig. \ref{fig:extragalactic}, we show the X-ray fluxes ($0.2-10$ keV) generated by the passage of a $10^5 \Msun$ IMBH in the five ISM environments for a range of distances between 1 kpc and 10 Mpc. We indicate the distance to a few example locations, from the Galactic center to the Andromeda galaxy, noting that within a radius of $4$ Mpc, there are more than 200 galaxies, although many of them are dwarfs \citep{galaxies_cat}.
We calculated the flux reported in Fig. \ref{fig:extragalactic} for the five ISM environments from their \textit{median luminosity}. As some environments exhibit a range of luminosities spanning $\sim 13$ orders of magnitudes (see Fig. \ref{fig:fluxes_selection}), the typical values of fluxes in Fig. \ref{fig:extragalactic} are indicative.

From Fig. \ref{fig:extragalactic}, we see that the median luminosities generated in the WNM, WIM, and HIM are invisible to AXIS or detectable only within the MW. On the contrary, fluxes generated in the CNM and MCs are detectable by an AXIS deep field well outside the MW. AXIS imaging reaching a depth of $\sim 3 \times 10^{-18} \rm \, erg \, s^{-1} \, cm^{-2}$ could detect the electromagnetic signature of the passage of an IMBH of $10^5 \Msun$ in hundreds of galaxies within $10$ Mpc distance.
Although MCs and CNM environments occupy only a small volume fraction of a typical MW-like galaxy ($0.05\%$ and $1\%$, respectively, see \cite{Ferriere_2001}), the availability of a large number of external galaxies within reach dramatically expands the chances of detecting such signatures.

As most of the X-ray luminosities of the sources considered here are $< 10^{40} \rm \, erg \, s^{-1}$, contamination from X-ray binaries (XRB) is of concern. To disentangle their emission, synergies with observatories in other wavelengths (e.g., JWST, Roman, and Rubin) will be fundamental.

%%%%%%%%%%%%%%%%%%%%%%%%%%%%%%%%%%%%%%%%%%
\section{Concluding Remarks}
To conclude, AXIS represents a significant improvement over current X-ray facilities and opens the way to detect a completely unknown population of black holes. In the MW and, even more crucially, in $>200$ local galaxies, AXIS can detect the X-rays emitted by the passage of IMBHs within dense ISM environments. Such detections are crucial for understanding the demographics and evolution of IMBHs and the merging history of galaxies.

%\vspace{6pt} 

%%%%%%%%%%%%%%%%%%%%%%%%%%%%%%%%%%%%%%%%%%

\acknowledgments{F.P. acknowledges support from a Clay Fellowship administered by the Smithsonian Astrophysical Observatory. This work was also supported by the Black Hole Initiative at Harvard University, which is funded by grants from the John Templeton Foundation and the Gordon and Betty Moore Foundation.
The authors kindly acknowledge the AXIS team for their outstanding scientific and technical work over the past year. This work results from several months of discussion in the AXIS-AGN SWG.
}

\section*{References}
\vspace{-30pt}
%=====================================
% References, variant B: external bibliography
%=====================================
\externalbibliography{yes}
\bibliography{references}

%%%%%%%%%%%%%%%%%%%%%%%%%%%%%%%%%%%%%%%%%%
\end{document}